\documentclass[aps,prl,epsfigure,showpacs,twocolumn]{revtex4}
\usepackage{dcolumn}
\usepackage{bm}
\usepackage{graphicx}
\usepackage{amsmath}
\usepackage{latexsym}
\usepackage{amsfonts}
\usepackage{amssymb}
\usepackage{array}
\usepackage{epsfig}

\newcommand{\ket}[1]{\left\vert#1\right\rangle}

\newcommand{\bra}[1]{\left\langle#1\right\vert}

\begin{document}

\title{Perfect state transfer on a spin-chain without state initialization}

\author{C. Di Franco, M. Paternostro, and M. S. Kim}

\affiliation{School of Mathematics and Physics, Queen's University, Belfast BT7 1NN, United Kingdom}

\date{\today}

\begin{abstract}
We demonstrate that perfect state transfer can be achieved using an engineered spin chain and clean local end-chain operations, without requiring the initialization of the state of the medium nor fine tuning of control-pulses. This considerably relaxes the prerequisites for obtaining reliable transfer of quantum information across interacting-spin systems. Moreover, it allows us to shed light on the interplay among purity, entanglement and operations on a class of many-body systems potentially useful for quantum information processing tasks.
\end{abstract}


\maketitle

The ability to prepare a fiducial state of a quantum system that has to accomplish a task of quantum communication or computation is one of seven {desiderata}, more commonly known as {\it DiVincenzo's criteria}~\cite{divincenzo}, that any reliable device for Quantum Information Processing (QIP) should meet. However, even the innocent request for a pure reference-state for the initialization of a QIP device is not easily granted, in practice, mainly due to the difficulty of preparing pure states of multipartite systems. A striking example is given by nuclear-magnetic-resonance QIP~\cite{nmr}, where the signal observed in an experiment comes from a chaotic ensemble of emitters, whose overall state is strongly mixed, and is ``reinterpreted'' quantum mechanically by relying on the concept of pseudo-purity~\cite{nmr2}. Another very important instance is provided by schemes for quantum state transfer (QST) in spin chains~\cite{unmodulated}. These have emerged as remarkable candidates for the realization of faithful short-distance transmission of quantum information~\cite{unmodulated2}. Although the preparation of the spin-medium in a fiducial pure state is an important step in the achievement of optimal transport fidelity, studies conducted on the effects of randomization of the chain's state have revealed that the process' efficiency gets spoiled, in a way that quantitatively depends on the mechanism assumed for such randomization~\cite{josephsonchain,rand}. 

Here we show that the conditions about the initial state of a spin chain which enable perfect state transfer can be considerably reduced, without requiring fine tuning of control-pulses over the chain~\cite{fitz}. Specifically, we demonstrate a scheme for perfect QST that is able to bypass the initialization of the spin-medium in a {\it known pure} state. The scheme requires only end-chain single-qubit operations and a {\it single} application of a global unitary evolution and is thus fully within a scenario where the control over the core part of the spin medium is relaxed in favour of controllability of the first and last element of the chain. We show flexibility of our designed protocol, which can be adapted to various interaction models. In fact, {\it we give the general conditions necessary in order to achieve perfect state transfer without state initialization via our scheme}. With minimal changes, one can use any Hamiltonian satisfying particular conditions on the time-evolution of two-site operators, as clearly identified in this paper. 

In order to provide the details of our protocol, we address the cases of two models for QST in spin chains which have been widely used so far~\cite{cambridge,josephsonchain}. We start with a nearest-neighbor Ising coupling involving $N$ spin-$1/2$ particles that also experience a transverse magnetic field. Its Hamiltonian reads $\hat{{\cal H}}_1=\sum^{N-1}_{i=1}J_{i}\hat{Z}_{i}\hat{Z}_{i+1}+\sum^N_{i=1}B_i\hat{X}_i$. Here, $J_i$ is the interaction strength between spin $i$ and $i+1$ and $B_{i}$ is the strength of the coupling of spin $i$ to a local magnetic field. In our notation, $\hat{X},\,\hat{Y}$ and $\hat{Z}$ denote the $x,\,y$ and $z$ Pauli matrix, respectively. We choose $J_i=J\sqrt{4i(N-i)}$ and $B_i=J\sqrt{(2i-1)(2N-2i+1)}$ with $J$ being a characteristic energy scale that depends on the specific physical implementation of the model (we choose units such that $\hbar=1$ throughout the paper). By applying the single-spin operation $(\hat{\openone}+i\hat{X})/\sqrt 2$ on the first element of the chain and using the eigenstates $\ket{\pm}=(\ket{0}\pm\ket{1})/\sqrt{2}$ of the $\hat{X}$ operator as computational basis, end-to-end perfect QST is achieved via $\hat{{\cal H}}_1$, when $Jt^*=\pi/4$ ($t^*$ is the evolution time) and for the initial fiducial state $\ket{+...+}_{2,...,N}$ of the spin-medium~\cite{josephsonchain}. This result has been obtained by analyzing the system from an information-flux (IF) viewpoint~\cite{informationflux}. However, the state-transfer fidelity is sensitive to deviations of the initial state from the one being ideally required.

For the understanding of the following discussion, it is enough to mention that the IF is in general rather useful when information regarding multi-site correlation functions is needed~\cite{matryoshka}. The analysis is performed in Heisenberg picture and requires $\hat{{\cal O}}(t)$, {\it i.e.} the time-evolved form of a given chain-operator $\hat{O}$. 
Here, for the purposes of our study, we concentrate on the evolution of two-site operators $\hat{\openone}_i\hat{X}_{N-i+1}$, $\hat{Z}_i\hat{Y}_{N-i+1}$, and $\hat{Z}_i\hat{Z}_{N-i+1}$. At time $t^*=\pi/4J$, by solving the relevant Heisenberg equations, we have that
\begin{equation}
 \begin{split}
  &\hat{\cal \openone}_i(t^*)\hat{\cal X}_{N-i+1}(t^*)=\hat X_i\hat \openone_{N-i+1},\\
  &\hat{\cal Z}_i(t^*)\hat{\cal Y}_{N-i+1}(t^*)=\hat Y_i\hat Z_{N-i+1},\\
  &\hat{\cal Z}_i(t^*)\hat{\cal Z}_{N-i+1}(t^*)=\hat Z_i\hat Z_{N-i+1}.
 \end{split}
\label{eq:evolution}
\end{equation}
Clearly, each of these two-site operators evolves in its swapped version, without any dependence on other chain's operators. This paves the way to the core of our protocol, which we now describe qualitatively. Qubit $1$ is initialized in the input state $\rho^{in}$ (either a pure or mixed state) we want to transfer and qubit $N$ is projected onto an eigenstate of $\hat{Z}$. Then the interaction encompassed by $\hat{\cal H}_1$ is switched on for a time $t^*=\pi/4J$, after which we end up with an entangled state of the chain. The amount of entanglement shared by the elements of the chain depends critically on their initial state, as it is commented later on. Regardless of the amount of entanglement being set, a $\hat{Z}$-measurement over the first spin projects the $N$-th one onto a state that is locally-equivalent to $\rho^{in}$. More specifically, if the product of the measurement outcomes at $1$ (after the evolution) and $N$ (before the evolution) is $+1$ ($-1$), the last spin will be in $\rho^{in}$ ($\hat{X}\rho^{in}\hat{X}$). In any case, apart from a simple single-spin transformation, perfect state transfer is achieved. For completeness of presentation, here we quantitatively assess the performance of our proposal. 

We start by considering spins $2,...,N-1$ all prepared in (unknown) eigenstates of the $\hat{Z}$ operator. For simplicity, we assume a pure state $\ket{\psi}=\alpha\ket{0}+\beta\ket{1}$ to be transmitted and the last spin in $\ket{0}$, although the generalization is straightforward. For definiteness, a representative of the initial state of the medium is written as $\ket{a_2...a_{N-1}}_{2,...,N-1}$ with $\ket{a_i}_i$ the state of spin $i$ ($a_i=0,1$). The final state of the chain, $e^{-i\hat{\cal H}_1t^*}\ket{\psi}_1\ket{a_2...a_{N-1}}_{2,...,N-1}\ket{0}_N$, is $\ket{\Psi}_{F}=(1/\sqrt{2})\,[\ket{0}_1\ket{a_{N-1}...a_2}_{2,...,N-1}\ket{\psi}_N+i\ket{1}_1\ket{a^\perp_{N-1}...a^\perp_2}_{2,...,N-1}(\hat{X}\ket{\psi}_N)]$, where $\bra{a^\perp_i}{a_i}\rangle=0,~\forall{i}$. Thus, upon measurement of the first spin over the $\hat{Z}$ eigenbasis, the state of the last spin is clearly locally-equivalent to $\ket{\psi}$ (and separable with respect to the subsystem $\{2,...,N-1\}$). The form of $\ket{\Psi}_{F}$ reveals the core of our mechanism. In fact, before the measurement stage, a fraction of genuine $N$-party entanglement of Greenberger-Horne-Zeilinger (GHZ) form~\cite{ghz} is shared by the elements of the chain. Such fraction is maximum for $\langle{\psi}|{\hat{X}}|{\psi}\rangle=0$ and disappears if $\ket{\psi}$ is taken as an eigenstate of $\hat{X}$, showing that the state to be transmitted acts as a knob for the entanglement in the chain. This consideration can be extended to any other spin of the medium. Indeed, suppose that one of the central spins (labelled $j$) is prepared in an eigenstate $\ket{\pm}_j$ of $\hat{X}$. The final state of the chain after the evolution driven by $\hat{\cal H}_1$ is $(1/\sqrt{2})\,\ket{\pm}_{N-j+1}[\ket{0}_1\ket{a_{N-1}...a_2}_{(2,...,N-1)'}\ket{\psi}_N+i\ket{1}_1\ket{a^\perp_{N-1}...a^\perp_2}_{(2,...,N-1)'}(\hat{X}\ket{\psi}_N)]$, where $(2,...,N-1)'$ denotes the set of all spins from $2$ to $N-1$, spin $N-j+1$ excluded. This shows that, in general, the GHZ entanglement shared by the elements of the chain before the measurement stage will not include the spins that are mirror-symmetrical with respect to any element initially prepared in an eigenstate of $\hat{X}$.

We can now extend our analysis to the case of an initial mixed state of the spin-medium. As before, for simplicity, the state of the last spin is $\ket{0}$. By following the same steps as above, the final state of the system would be given by $\rho_F=[\ket{0}_1\!\bra{0}\otimes\rho\otimes\rho^{in}_N\!+\!\ket{1}_1\!\bra{1}\otimes\hat{\cal S}_2\rho\otimes\hat{\cal T}_2\rho^{in}_N-(i\ket{0}_1\!\bra{1}\otimes\hat{\cal S}_1\rho\otimes\hat{\cal T}_1\rho^{in}_N\!+h.c.)]/2$ with $\rho$ the density matrix of spins from $2$ to $N-1$ obtained by applying a mirror-inversion operation on their initial state and $\rho^{in}$ the density matrix of the state to transfer. We have defined
$\hat{\cal S}_1\rho=\rho\prod_{i=2}^{N-1}\hat{X}_i$, $\hat{\cal S}_2\rho=\prod_{i=2}^{N-1}\hat{X}_i\rho\prod_{i=2}^{N-1}\hat{X}_i$, $\hat{\cal T}_1\rho^{in}_N=\rho^{in}_N\hat{X}_N$, $\hat{\cal T}_2\rho^{in}_N=\hat{X}_N\rho^{in}_N\hat{X}_N$. Again, the crucial point here is that, regardless of the amount of entanglement established between the spin-medium and the extremal elements of the chain ({\it i.e.} spins $1$ and $N$), upon $\hat{Z}$-measurement of $1$, the last spin is disconnected from the rest of the system, {\it whose initial state is inessential to the performance of the protocol} and could well be, for instance, a thermal state of the chain in equilibrium at finite temperature. In fact, the key requirements for our scheme are the arrangement of the proper time-evolution (to be accomplished within the coherence times of the system) and the performance of clean projective measurements on spin $1$ and, preventively, on $N$.

The last requirement of our scheme is particularly important and, in order to estimate its relevance, we evaluate the performance of the protocol against the purity of the initial state of spin $N$. For the sake of simplicity, we focus on the case in which a pure state is transmitted, the case of mixed states being easily deduced.  Our instrument is the input-output transfer fidelity $F_{\text{transfer}}=\bra{\psi}\rho^{out}_N\ket{\psi}$ (with $\rho^{out}_N$ the state of the last qubit after the protocol), which is unity when the two states are the same and zero when they are mutually orthogonal. We have that $F_{\text{transfer}}=p_{00}+(1-p_{00})\text{Tr}(\ket{\psi}\!\bra{\psi}{\hat X}),$ where $p_{00}\in[0,1]$ is the population of $\ket{0}$ in the density matrix describing the initial state of spin $N$ (decomposed over the $\hat{Z}$-basis). The independence of the state fidelity from the coherences of the initial state of $N$ implies that it is effectively the same to operate with a pure state $\ket{\phi}_N=\sqrt{p_{00}}\ket{0}_N+e^{i\varphi}\sqrt{1-p_{00}}\ket{1}_N$ or the mixed one $\rho_N=p_{00}\ket{0}\bra{0}+\gamma\ket{0}\bra{1}+\gamma^*\ket{1}\bra{0}+(1-p_{00})\ket{1}\bra{1}$, with $\gamma$ being arbitrary. Any error in the QST process has to be ascribed to the fact that, for a non-unit value of $p_{00}$, the perfectly-transmitted state $\ket{\psi}$ has an admixture with the ``wrong'' state $\hat{X}\!\ket{\psi}$. This explains the dependence of $F_{\text{transfer}}$ on the state to be transmitted (more precisely, on $\langle{\psi}|{\hat{X}}|{\psi}\rangle$). 

\begin{figure}
\psfig{figure=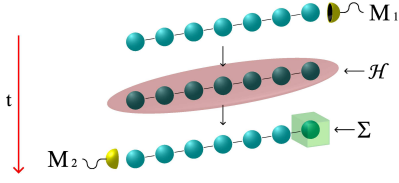,height=2.5cm}
\caption{Sketch of the scheme for perfect QST. $M_1$ and $M_2$ are measurements performed over a fixed basis, $\Sigma$ is a conditional operation, and $\hat{\cal H}$ is the Hamiltonian.} 
\label{protocols}
\end{figure}

As anticipated, our results are not bound to the specific instance of interaction model being considered but, more generally, on the way two-site operators evolve in time. Under different couplings, similar behaviors for objects like $\hat{\cal O}_i(t^*)\hat{\cal O}_{N-i+1}(t^*)$ can be observed, therefore leading to conclusions similar to those put forward in our discussion so far. In fact, with rather minor adjustments to the procedure described above, one can apply the scheme to $N$ spin-$1/2$ particles coupled via the $XX$ model $\hat{{\cal H}}_2=\sum^{N-1}_{i=1}K_{i}(\hat{X}_{i}\hat{X}_{i+1}+\hat{Y}_i\hat{Y}_{i+1})$ with $K_{i}=J\sqrt{i(N-i)}$. $\hat{\cal H}_2$ has been extensively analyzed~\cite{cambridge}: $1\rightarrow{N}$ perfect QST is achieved when the initial state of all the spins but the first one is $\ket{0}$. However, let us reason in terms of IF again, proceed as done above for the Ising model and look at the dynamics of two-site operators symmetrical with respect to the center of the chain. At time $t^*=\pi/4J$, we have that, for any $N$, $\hat{\cal \openone}_i(t^*)\hat{\cal Z}_{N-i+1}(t^*)=\hat Z_i\hat \openone_{N-i+1}$. On the other hand, for even $N$ we find
\begin{equation}
 \begin{split}
  &\hat{\cal X}_i(t^*)\hat{\cal X}_{N-i+1}(t^*)=\hat X_i\hat X_{N-i+1},\\
  &\hat{\cal X}_i(t^*)\hat{\cal Y}_{N-i+1}(t^*)=\hat Y_i\hat X_{N-i+1},
 \end{split}
\label{eq:evolutioncambridge}
\end{equation}
while for an odd number of spins in the chain we have
\begin{equation}
 \begin{split}
  &\hat{\cal X}_i(t^*)\hat{\cal X}_{N-i+1}(t^*)=\hat Y_i\hat Y_{N-i+1},\\
  &\hat{\cal X}_i(t^*)\hat{\cal Y}_{N-i+1}(t^*)=\hat X_i\hat Y_{N-i+1}.
 \end{split}
\label{eq:evolutioncambridge2}
\end{equation}
The procedure to follow has to be adjusted depending on the chain's length. In particular, the last spin has to be projected onto $\ket{\pm_N}=(\ket{0}\pm e^{iN\frac{\pi}{2}}\ket{1})/\sqrt{2}$. In what follows, we say that outcome $+1$ ($-1$) is found if a projection onto $\ket{+_N}$ ($\ket{-_N}$) is performed. This change of basis with respect to the protocol designed for the Ising model is due to the different form of the {\it transverse} nature of $\hat{\cal H}_2$. After the evolution $e^{-i\hat{\cal H}_2t}$, we measure the first spin over the $\hat{X}$ eigenbasis. The resulting output state depends on the product of the measurement outcomes at $1$ (after the evolution) and $N$ (before the evolution). If such product is $+1$ ($-1$), the transmitted state will be $(\hat{T}^N)^\dag\rho^{in}(\hat{T}^N)$ [$(\hat{T}^N)\rho^{in}(\hat{T}^N)^\dag$], where $\hat{T}=\ket{0}\bra{0}+e^{i\frac{\pi}{2}}\ket{1}\bra{1}$ (therefore, $\hat{T}^2=\hat{Z}$). Also in this case, apart from a single-spin transformation, perfect state transfer is achieved. A sketch of the general scheme for perfect state transfer is presented in Fig.~\ref{protocols}.
For the XX model, we can write the final state of the system before the $M_2$-measurement stage (we consider the last spin in $\ket{+_N}$), as $\rho_F=\{\ket{+}_1\!\bra{+}\otimes\tilde{\rho}\otimes\tilde{\rho}^{in}_N\!+\!\ket{-}_1\!\bra{-}\otimes\hat{\cal S}_4\tilde{\rho}\otimes\hat{\cal T}_4\tilde{\rho}^{in}_N-[i(-1)^{N}\ket{+}_1\!\bra{-}\otimes\hat{\cal S}_3\tilde{\rho}\otimes\hat{\cal T}_3\tilde{\rho}^{in}_N+h.c.]\}/2$ with $\tilde{\rho}\!=\!\hat{A}^\dag\rho\hat{A}$, $\tilde{\rho}^{in}\!=\!(\hat{T}^N)^\dag\rho^{in}(\hat{T}^N)$, $\hat{A}=\prod_{i=2}^{N-1}\hat{T}_i^N$, $\hat{\cal S}_3\tilde{\rho}=\tilde{\rho}\prod_{i=2}^{N-1}\hat{Z}_i$, $\hat{\cal S}_4\tilde{\rho}=\prod_{i=2}^{N-1}\hat{Z}_i\tilde{\rho}\prod_{i=2}^{N-1}\hat{Z}_i$, $\hat{\cal T}_3\tilde{\rho}^{in}_N=\tilde{\rho}^{in}_N\hat{Z}_N$, $\hat{\cal T}_4\tilde{\rho}^{in}_N=\hat{Z}_N\tilde{\rho}^{in}_N\hat{Z}_N$.

In general, the protocol can be adapted to any Hamiltonian for which we can find a triplet of single-spin operators $\hat{\cal B},\hat{C},\hat{D}$ such that, for symmetric spin pairs, we have $\hat{\cal B}^{j_{O}}_i(t^*)\hat{C}_{N-i+1}\hat{\cal O}_{N-i+1}(t^*)=O_{i}\hat{D}^{k_O}_{N-i+1}$. Here, $\hat{\cal B}_{i}$ ($\hat{D}_{N-i+1}$) provides the eigenbasis for the measurement over spin $i$ ($N-i+1$) of the chain after (before) the evolution, $\hat{C}_{N-i+1}$ is a decoding operation, $\hat{O}_i=\hat{\cal O}_i(0)=X,Y,Z$ and $j_O,k_O=0,1$, depending on the coupling model. For instance, Eq.~(\ref{eq:evolution}) are gained by taking $\hat{\cal B}_i=\hat{\cal Z}_i$, $\hat{C}_{N-i+1}=\openone$, $\hat{D}_{N-i+1}=\hat{Z}_{N-i+1}$ with $j_X=k_X=0,j_{Y,Z}=k_{Y,Z}=1$. We point out that when these conditions are not fulfilled, as in Ref.~\cite{campos} where an antiferromagnetic Heisenberg chain is used, our protocol can still be rather successful. In these cases, through IF we can calculate an estimate of the average transfer fidelity~\cite{informationflux}. For instance, for a homogeneous XX model of $N=100$ spins with end-point coupling strengths $J_{1,N-1}$ such that $J_{1}=J_{N-1}\simeq{0.7}J$, the average transfer fidelity via our protocol is estimated to be $\ge0.87$.

As noticed for the case of $\hat{\cal H}_1$, the nature and amount of the entanglement generated during the performance of the protocol depends on the form of the initial state of the medium's spins. Multipartite entanglement shared by all (or some of) the elements of the chain, as well as only bipartite entanglement involving first and last qubits (as in the case when spins from $2$ to $N-1$ are in eigenstates of $\hat{X}$) can be generated. Nevertheless, unit fidelity of transfer is achieved when the right time evolution and perfect hard-projections are in order. This strongly supports the idea that QST protocols do not crucially rely on the specific nature and quality of the entanglement generated throughout the many-body dynamics, in stark contrast with other schemes for QIP~\cite{grover}.

On the other hand, the counterintuitive fact that $F_{\text{transfer}}=1$ regardless the initial state of medium could remind one, at first sight, the idea of {\it deterministic quantum computation with one quantum bit} (DQC1) proposed in Ref.~\cite{knill}. In this model, a single pure two-level system and arbitrarily many ancillae prepared in a maximally mixed state are used in order to solve problems for which no efficient classical algorithm is known. The apparent similarity with our case is resolved by observing that  in DQC1 the initial state is restricted to that particular instance (a pure single-qubit state and a maximally mixed state of all the other qubits), which can be seen as the ``fiducial'' state invoked in DiVincenzo's criterion. Differently, our scheme completely relaxes the knowledge required about the state of the spin-medium, which might well be completely unknown to the agents that perform the QST process. The achievement of quantum computation with initial mixed states has also been analyzed in Ref.~\cite{parker}, where it has been shown that a single qubit supported by a collection of qubits in an arbitrary mixed state is sufficient to efficiently implement Shor's factorization algorithm. In this case, however, the performance of the protocol depends on the input state. Indeed, the average efficiency over all the possible random states (mixed or pure) is evaluated, but for some particular input states (for instance, $\ket{0...0}_{2,...,N}$) it can drop below classical limit.
Differently, our scheme is independent of the initialization of the spin-medium and its efficiency cannot be spoiled by any input state.

It is worth clarifying an important aspect which is certainly apparent to the careful reader. The procedures described so far might remind one of the general scheme for one-way quantum computation put forward in Ref.~\cite{OW}. In both cases, the optimal result of a protocol depends on the performance of perfect projective measurements onto specific elements of a register and the feed-forward of a certain amount of classical information (in our case, the outcome of the measurements over spin $1$ and/or the initial projection of spin $N$). Moreover, as in the one-way model, in our proposal the ``pattern'' of quantum correlations depends on the initial state of the elements of the system. However, such an analogy cannot be pushed too forward as, remarkably, the use of quantum entanglement in the two protocols is different. While the one-way model relies on a pre-built multipartite entangled resource (the graph state) which is progressively destroyed by a proper program of measurements, in our scheme the multipartite entanglement (if any) is built while the protocol is running. We just need a single measurement for the processing of the information encoded at the input state. In addition, differently from a graph state, the preparation of some of the spins in the medium in states preventing their participation to a multipartite entangled state does not spoil the efficiency of the protocol, as we have demonstrated. This is not the case for a graph-state built out of pairwise Ising interactions: the wrong initialization of a part of the register excludes it from the overall entangled state, and actually ``blocks'' the transfer of information through that region of the register.

Finally, we would like to stress the difference between our opproach and those achieving perfect QST via mirror-inverting coupling model~\cite{mirror}. In our general protocol, mirror inversion is ``induced'' in models which otherwise would not allow it, by adjusting the pattern of quantum interferences within the spin medium via the encoding/decoding local stages. By means of these, one can avoid the pre-engineered fulfilment of precise conditions on the spectrum of each interacting spin~\cite{mirror,yung} which, combined with reflection symmetry, are required for mirror-inversion. Our models satisfy just the second of these conditions, perfect QST without initialization being achieved through the encoding and decoding steps we have described. 

We have shown the existence of a simple control-limited scheme for the achievement of perfect QST in a system of interacting spins without the necessity of demanding state initialization. Our flexible protocol requires just {\it one-shot} unitary evolution and end-chain local operations. Its efficiency arises from the establishment of {\it correlations} between the first and last spin of the transmission-chain. With the exception of limiting cases where the transfer is automatically achieved [as for the transfer of eigenstates of $\hat{X}_1$ ($\hat{Z}_1$) when model $\hat{\cal H}_1$ ($\hat{\cal H}_2$) is used], these are set regardless of the state of the spin medium, their amount being a case-dependent issue. The end-chain measurements, which are key to our scheme, ``adjust'' such correlations in a way so as to achieve perfect QST. We hope that our study, which paves the way to a thorough investigation about the role played by multipartite entanglement in perfect QST, would help in the experimental realization of short-distance quantum communication in, for instance, engineered superconducting chains or patterned distributed nanosystems.

C.D.F. thanks D. Ballester and G. A. Paz-Silva for discussions. We acknowledge support from UK EPSRC and QIPIRC. M.P. is supported by the EPSRC (EP/G004579/1).

\end{document}